\documentclass[aps,showpacs,superscriptaddress,longbibliography]{revtex4-1}
\usepackage{amsfonts}
\usepackage{amssymb}
\usepackage{amsmath}
\usepackage{graphicx,color}
\usepackage{bm}
\usepackage{ulem}
\usepackage{physics}
\usepackage{siunitx}
\usepackage{lineno}

\begin{document}
%低エネルギーではビーム損失が主になる領域では有効でないことの考慮と説明の追加（計算はされているが陽に議論する方が良いかも）
%
%\linenumbers

\title{Breakeven Conditions for Beam–Target Fusion with Electron‑Suppressed Targets}
\date{\today}

\author{Tadafumi Kishimoto}
\affiliation{Research Center for Nuclear Physics, Osaka University, Ibaraki, Osaka 567-0047, Japan}

\begin{abstract}
This manuscript provides a detailed and extended analysis of the breakeven conditions for nuclear fusion based on beam–target interactions, distinct from conventional plasma-based approaches.  Building on the energy-based criterion introduced in the accompanying Letter~\cite{TK_arXiv}, we formulate a self-consistent description of stopping power in electron-suppressed targets and derive quantitative, implementation-agnostic conditions under which fusion energy generation can exceed beam energy loss.
\end{abstract}

\maketitle

\section{Introduction}
Achieving breakeven—where fusion output equals or exceeds the input—is a universal benchmark in fusion research and is often framed in terms of the Lawson criterion~\cite{lawson1955, Wurzel2022Lawson}. 
In this work, we focus on beam–target interactions and examine under what conditions breakeven can be realized in that setting, distinct from conventional plasma-based approaches. 

Building on the accompanying Letter~\cite{TK_arXiv}, which demonstrated that the energy generated by nuclear reactions can exceed energy loss under appropriate conditions, the present paper provides a self-contained and implementation-agnostic formulation to assess that statement quantitatively and to identify the corresponding feasibility window—i.e., the parameter regime in which generation surpasses loss and the breakeven condition is satisfied.

We develop a physically grounded treatment of energy loss specific to electron-suppressed targets (electron-free in the ideal limit), making explicit how the relevant long- and short-distance scales bound the logarithmic factor that controls stopping power. We further clarify which processes contribute as continuous energy loss and which act as absorptive channels that remove beam flux. The detailed formulation is presented in the main text.

For context, recent inertial-confinement experiments at the National Ignition Facility (NIF) have demonstrated so-called scientific breakeven (target gain $>1$), in which the fusion energy released from the fuel exceeds the laser energy coupled to it~\cite{NIF_PRL_2024_gain}. In contrast, the present work addresses a beam–target route and treats breakeven explicitly as generation exceeding loss within a stopping framework, independent of facility-level efficiencies. The two notions are complementary but conceptually distinct.

The remainder of the paper is organized as follows. We first introduce the local and integrated criteria used to compare fusion energy generation with beam energy loss. We then present the modeling assumptions and their associated physical cutoffs. Finally, we quantify the implications for representative fusion reactions and discuss their relevance for experimentally achievable conditions.

\section{Standard Nuclear Fusion Reactions and Breakeven Conditions}
Typical nuclear fusion reactions considered in energy research include:
\begin{itemize}
  \item D + T $\rightarrow$ $\alpha$ + n \hfill ($Q = 17.59~\mathrm{MeV}$)
  \item D + D $\rightarrow$ T + p \hfill ($Q = 18.35~\mathrm{MeV}$)
  \item D + D $\rightarrow$ $^3$He + n \hfill ($Q = 11.32~\mathrm{MeV}$)
  \item D + $^3$He $\rightarrow$ $\alpha$ + p \hfill ($Q = 18.35~\mathrm{MeV}$)
  \item p + $^{11}$B $\rightarrow$ 3$\alpha$ \hfill ($Q = 11.1~\mathrm{MeV}$)
\end{itemize}
All of these reactions are exothermic. 
Among these possibilities, we focus on the deuterium--tritium (DT) reaction in the following, as it provides the most favorable testing ground for beam--target breakeven at low center-of-mass energies owing to its relatively large fusion cross section.

\begin{figure}[h]
\centering
\includegraphics[width=0.6\textwidth]{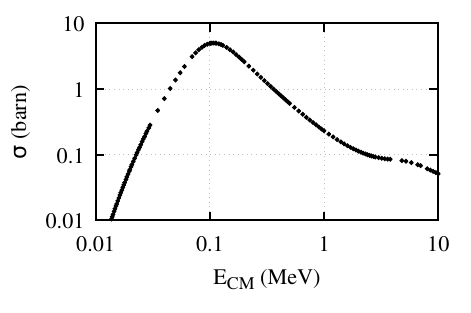}
\caption{Deuterium--tritium (DT) fusion cross section as a function of the center-of-mass energy.~\cite{DT_sigma}}
\label{fig:dt_cross_section}
\end{figure}

Figure~\ref{fig:dt_cross_section} shows a representative evaluation of the DT fusion cross section as a function of the center-of-mass energy. 
The pronounced enhancement of the cross section in the sub-MeV range highlights why DT is particularly well suited for probing breakeven conditions in beam--target configurations, which are the primary focus of the present work.

\section{Energy Loss Mechanisms and Breakeven Conditions}

We examine breakeven in a beam--target configuration, in which a deuteron beam of laboratory energy $E$ impinges on a stationary tritium target. In the present formulation, the differential target thickness $dx$ is treated as a length element, so that the product $\rho\,dx$ represents the areal target density (in units of mol\,cm$^{-2}$). The fusion energy generated per areal thickness is

\begin{equation}
  \frac{dE_{\mathrm{gen}}}{dx} = \rho\,\sigma(E)\,Q,
  \label{eq:gen}
\end{equation}
where $\rho$ is the target number density expressed in mol\,cm$^{-3}$, $\sigma(E)$ the fusion cross section, and $Q$ the reaction energy release. 

The energy loss per areal thickness is given by the stopping power, for which we adopt a Bethe–Bloch–type expression
\begin{equation}
\frac{dE_{\mathrm loss}}{dx}
= K\,\frac{1}{m_e\,\beta^2}\,
\ln\!\left(\frac{T_{\max}}{T_{\min}}\right),
\label{eq:E_lossBB}
\end{equation}
where $\beta$ is the projectile speed, $m_e$ the electron mass, and $K$ collects material- and unit-dependent constants.
The physically consistent cutoffs $T_{\max}$ and $T_{\min}$ (or equivalently $b_{\min},b_{\max}$) are discussed in Sec.~V.

A convenient local (differential) measure of breakeven is the ratio
\begin{equation}
%  R(E) \equiv \frac{(dE_{\mathrm{gen}}/dx)}{(dE/dx)_{\mathrm{loss}}},
 R(E) = \dv{E_{\mathrm{gen}}}{x} \,\bigg/\, \dv{E_{\mathrm{loss}}}{x},
  \label{eq:R_local_intro}
\end{equation}
so that the local breakeven condition is $R(E)>1$. In the limit of vanishing stopping power, a projectile would maintain its energy and eventually undergo fusion with unit probability, continuously generating fusion energy. Conversely, a sufficiently large stopping power halts the projectile before a reaction can occur, suppressing net energy gain. The boundary between these regimes is therefore defined by $R(E)=1$, which corresponds to the equality between the externally supplied energy required to compensate stopping losses and the fusion-generated energy.

When $R(E)>1$ holds over a finite energy interval along the slowing-down trajectory, the cumulative (integrated) gain becomes positive. Section~VI formalizes this condition and the following sections present the corresponding numerical results.   

%% INSERTED (DeltaE from momentum transfer)

\section{Stopping as a limiting factor}
\label{sec:stopping-role}
\begin{figure}[t]
  \centering
  \includegraphics[width=0.6\textwidth]{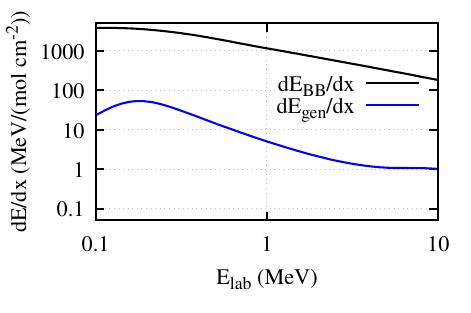}
  \caption{Differential comparison in conventional electron-containing materials: generation $dE_{\mathrm gen}/dx=\rho\,\sigma(E)\,Q$ and loss $dE_{\mathrm loss}/dx$ as functions of the laboratory beam energy. The black curve corresponds to the Bethe–Bloch stopping term, while the blue curve shows the fusion energy-generation term. The loss follows Eq.~\ref{eq:E_lossBB}, obtained from NIST stopping values for proton beams and converted to deuterons at equal projectile speed. Over the relevant energy range, the loss dominates, and the local index $R(E)$ remains at $\mathcal{O}(10^{-2})$.}
  \label{fig:E_lossBB_gene}
\end{figure}

Equation~\ref{eq:E_lossBB} illustrates that the stopping term enters Eq.~\ref{eq:R_local_intro} in a way that directly suppresses the local breakeven condition.
Based on the standard DT cross sections and the stopping model of Eq.~\ref{eq:E_lossBB}, we first evaluate, in a consistent manner, the differential energy \textit{generation} and \textit{loss} in conventional electron-containing materials. Figure~\ref{fig:E_lossBB_gene} presents $dE_{\mathrm gen}/dx=\rho\,\sigma(E)\,Q$ and $dE_{\mathrm loss}/dx$ as functions of the \textit{laboratory} beam energy. 

For the loss, we adopt NIST tabulated stopping values~\cite{nist_stopping_power} for proton beams in conventional matter and convert them to deuterons at the same projectile speed (equal-$\beta$ conversion), which is appropriate for Bethe--Bloch--type stopping where the dominant dependence enters through $\beta$ and a logarithmic factor. Over the relevant energy range, the loss exceeds the generation by about two orders of magnitude, corresponding to
\[
R(E)=\frac{dE_{\mathrm gen}/dx}{dE_{\mathrm loss}/dx}\sim\mathcal{O}(10^{-2}).
\]
As a consequence, nuclear reactions in conventional beam--target setups remain far from the local breakeven condition $R(E)=1$. In practice, this has historically motivated plasma-based approaches, in which high temperature, density, and confinement time are employed to compensate stopping losses.

We now examine the origin of the stopping term in order to assess whether there is any room to reduce it.
At the microscopic level, the energy transfer in a single small-angle Coulomb encounter is determined by the transferred momentum and the mass of the scatterer,
\begin{equation}
  \Delta E = \frac{(\Delta p)^2}{2m},
  \label{eq:single_collision_energy_transfer}
\end{equation}
within the nonrelativistic regime.
Here $\Delta p$ is set by the Coulomb force and hence by the charge and its spatial distribution; to leading order, it is independent of the scatterer mass. Consequently, for a given impulse $\Delta p$, the energy transfer scales inversely with $m$, implying that lighter scatterers accept larger energy transfers and therefore dominate stopping in conventional matter. This observation identifies the key lever for reducing the loss.

In the following sections, we establish the correspondence between the bounds on energy transfer and the impact-parameter cutoffs that control the logarithmic factor in Eq.~\ref{eq:E_lossBB}. 
This analysis naturally motivates the electron-free target concept introduced in the next section, after which concrete, physically consistent choices of the cutoffs are specified and their implications are explored.

\section{Electron-Free Target Concept}
\label{sec:ef-concept}

The analysis of the previous section shows that stopping in conventional materials is dominated by energy transfer to light electronic scatterers. Guided by Eq.~\ref{eq:single_collision_energy_transfer} and the fact that the transferred momentum $\Delta p$ is fixed by Coulomb forces, we therefore consider the idealized limit in which electronic scatterers are eliminated, and make explicit the correspondence that controls the logarithmic factor in Eq.~\ref{eq:E_lossBB}. 

The admissible impact-parameter interval $[\,b_{\min},\,b_{\max}\,]$ maps to the kinematic energy-transfer bounds via
\[
T_{\max}=T(b_{\min}),\qquad T_{\min}=T(b_{\max}),
\]
so that, in the small-angle regime,
\[
\ln\!\frac{T_{\max}}{T_{\min}} \;=\; 2\,\ln\!\frac{b_{\max}}{b_{\min}}.
\]
This relation makes explicit that the logarithmic factor in Eq.~\ref{eq:E_lossBB} is entirely fixed once the physical cutoffs
$\{b_{\min},b_{\max}\}$ (equivalently $\{T_{\max},T_{\min}\}$) are specified.

In electron-suppressed targets, $b_{\max}$ is set by geometric or collective screening of the beam--target system, 
while $b_{\min}$ is determined by the onset of processes that remove the projectile from the continuous-stopping channel.
Concrete, physically consistent choices of these cutoffs and their quantitative
consequences are discussed in the following sections.

\section{Impact-Parameter Cutoffs in Electron-Free Targets}
\label{sec:ef-bounds}

Building on the correspondence established in Sec.~\ref{sec:ef-concept}, we now quantitatively specify the impact-parameter cutoffs that determine the logarithmic
factor in Eq.~\ref{eq:E_lossBB} for electron-free targets. 
The logarithmic dependence in the Bethe--Bloch stopping term originates from integrating the small-angle Coulomb scattering contribution,
which scales as $1/b^{2}$, over the transverse phase space $2\pi b\,db$ between the impact-parameter cutoffs.

\subsection{Upper cutoff $b_{\max}$.}
\begin{itemize}
\item \textbf{Conventional (electron-containing) materials.}
Electrons are bound to atoms; unless ionized, they do not appear as free scatterers. Hence the smallest admissible energy transfer on the electronic side is set by the \textit{ionization energy}, and $b_{\max}$ is fixed by the value at which a Coulomb encounter just reaches that threshold, i.e.,
$T_{\min}=T(b_{\max})$. This reproduces the usual lower bound in $T$ that appears in standard tabulations of stopping powers.

\item \textbf{Electron-free targets.}
There is no bound-electron threshold, so $b$ is not limited by atomic ionization. Instead, the
effective upper bound is set by geometry and collective screening of the \textit{beam–target system}.
A single projectile experiences no net force from a spatially uniform charge distribution, so the
scale at which the beam’s ion distribution becomes effectively uniform bounds the useful $b$.
A natural and conservative choice is the \textit{mean inter-ion spacing} in the beam,
\[
b_{\max} \simeq \bar{d} \equiv n^{-1/3},
\]
with $n$ the ion number density seen by the projectile. In the estimates below we adopt $b_{\max}=300\,\mathrm{nm}$, corresponding to 
$n\simeq (3\times10^{-7}\,\mathrm{m})^{-3}\sim 4\times10^{19}\,\mathrm{m}^{-3}$, and broadly representative of $n\sim 10^{18}$--$10^{20}\,\mathrm{m}^{-3}$.
Because the logarithm in Eq.~(2) depends only on $b_{\max}$ through $\ln(b_{\max}/b_{\min})$, even a three-order-of-magnitude variation in $n$ 
(and thus in $b_{\max}$) changes the logarithmic factor at the level of order ten percent in the totals used here, and does not alter the qualitative conclusions.
\end{itemize}

We note that a more refined determination of $b_{\max}$—including explicit beam geometry and time structure—will be valuable in future work; owing to the logarithmic sensitivity in Eq.~(2), however, the conclusions presented here are robust against reasonable variations of $b_{\max}$.

\subsection{Lower cutoff $b_{\min}$ and the definition of $T_{\max}$}
\label{sec:bmin}

In the mapping of Eq.~(2), the lower impact-parameter cutoff $b_{\min}$ sets the upper energy-transfer bound via $T_{\max}=T\!\bigl(b_{\min}\bigr)$. 
Physically, $b_{\min}$ marks the boundary at which the description in terms of continuous small-angle Coulomb scattering (stopping power) ceases 
to be appropriate. Two classes of events are therefore removed from the continuous-stopping channel and counted as beam attenuation (flux loss) instead:
\begin{itemize}
  \item[(i)] large-angle Coulomb events that strongly decelerate or redirect the projectile in a single collision, and
  \item[(ii)] nuclear reactions in which the projectile is effectively removed or converted.
\end{itemize}

\paragraph*{Kinematic threshold for large-angle events (flux loss).}
To exclude large-angle Coulomb events from the continuous-stopping integral, we adopt a kinematic half-speed threshold: if a single collision reduces 
the projectile speed below one half of its incoming value, the subsequent stopping rises sharply (the canonical $1/\beta^{2}$ dependence), and
the particle is effectively removed from the useful beam. In the CM frame, the deflection angle $\theta$ and impact parameter $b$ are related through the Rutherford expression
\begin{equation}
  b(\theta;E) = \frac{\kappa}{2E}\,\cot\!\frac{\theta}{2}, \qquad
  \kappa \equiv Z_1 Z_2 e^2,
  \label{eq:rutherford_btheta}
\end{equation}
valid in the small-angle regime. The critical impact parameter for beam loss is then
\begin{equation}
  b_{\mathrm{crit}}(E) \equiv b(\theta_c;E),
  \label{eq:bcrit}
\end{equation}
where $\theta_c$ is fixed by the half-speed condition. Events with $\,b<b_{\mathrm{crit}}(E)\,$ are classified as flux loss and excluded from the continuous-stopping integral.
We note that the logarithmic dependence obtained from the $1/b^2$ form corresponds to a conservative estimate: at very small impact parameters the Coulomb interaction is naturally regularized, effectively replacing $1/b^2$ by $1/(b_0^2+b^2)$ and further suppressing the stopping contribution. Our treatment therefore provides an upper bound on the energy loss.

\paragraph*{Nuclear reactions as flux loss: geometric correspondence.}
Although reactions are quantum processes, the reaction cross section $\sigma_{\mathrm{reac}}(E)$ can be mapped to an effective area, $\pi b_{\mathrm{reac}}^2(E)\simeq \sigma_{\mathrm{reac}}(E)$, which sets the region in impact-parameter space where the projectile is lost by reaction. The domain
$\,b<b_{\mathrm{reac}}(E)\,$ is therefore also excluded from the continuous-stopping channel and counted as attenuation. 

Although the fusion cross section $\sigma$ is a quantum-mechanical quantity, its use for defining an effective geometrical radius $b_{\mathrm{reac}} = (\sigma/\pi)^{1/2}$ is well justified in the present semi-classical framework.  In the energy range considered here, the fusion process is dominated by low partial waves and can be effectively regarded as a short-range absorptive channel.  Reasonable variations of this prescription for $b_{\min}$ do not lead to noticeable changes in the calculated energy-loss spectra.

\paragraph*{Working definition of $b_{\min}$ and $T_{\max}$.}
Collecting the two exclusions, we define the lower cutoff for the continuous-stopping integral as
\begin{equation}
  b_{\min}(E) \;=\; \max\!\bigl\{\, b_{\mathrm{crit}}(E),\; b_{\mathrm{reac}}(E) \,\bigr\},
  \label{eq:bmin_definition}
\end{equation}
which translates to
\begin{equation}
  T_{\max}(E) \;=\; T\!\bigl(b_{\min}(E)\bigr).
  \label{eq:Tmax_from_bmin}
\end{equation}
With this convention, the stopping term in Eq.~(2) accounts exclusively for small-angle Coulomb scattering, while the beam-intensity decrease is described separately by an effective attenuation equation $\,dI/dx=-\sigma_{\mathrm{eff}}(E)\,I\,$ (see Sec.~\ref{sec:lab_to_cm}), where $\sigma_{\mathrm{eff}}$ collects the losses due to large-angle Coulomb events and nuclear reactions.

\paragraph*{Dominant contribution to $b_{\min}$.}
Over almost the entire beam-energy range considered here, the lower cutoff $b_{\min}(E)$ is determined by the reaction scale $b_{\mathrm{reac}}(E)$.
The kinematic cutoff for large-angle Coulomb scattering, $b_{\mathrm{crit}}(E)$, exceeds $b_{\mathrm{reac}}(E)$ only at very low beam
energies, $E \lesssim 0.1~\mathrm{MeV}$. In this energy region, however, the beam flux is already strongly attenuated
by nuclear reactions and large-angle scattering. Consequently, the contribution from this low-energy domain to the overall
energy loss is negligible, and $b_{\min}$ is effectively set by $b_{\mathrm{reac}}$ throughout the energy range relevant to the present analysis.

\begin{figure}[t]
  \centering
  \includegraphics[width=0.82\linewidth]{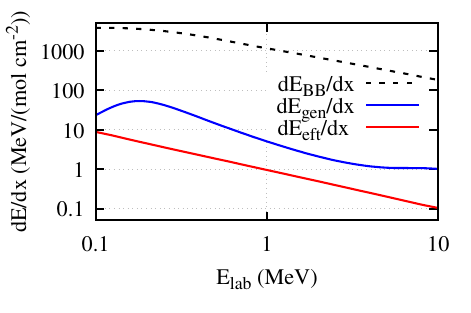}
  \caption{% 
Stopping power in an electron-free target as a function of the beam energy $E_{\mathrm{lab}}$.  The black dashed curve shows the Bethe--Bloch reference,
the blue curve the fusion energy-generation term, and the red curve the stopping power evaluated for the electron-free target.
The energy region where $b_{\mathrm{crit}}$ dominates over $b_{\mathrm{reac}}$
($E\lesssim0.12\,\mathrm{MeV}$) contributes negligibly once beam attenuation is taken into account.
  }
  \label{fig:E_lossBB_gene_losseft}
\end{figure}

\paragraph*{Energy loss for an electron-free target.}
We first note that the ratio $R(E)\equiv (dE_{\mathrm{gen}}/dx)/(dE_{\mathrm{loss}}/dx)$ defined in Eq.~(\ref{eq:R_local_intro})) satisfies $R(E)>1$ over the energy range relevant 
to this section. 
This condition implies that the maximum transferable energy is governed by the projectile--target mass kinematics rather than the electron case,
leading to an effective suppression factor of order $1/1000$--$1/2000$ instead of the electron-scattering value $m_e/M \simeq 1/5500$.
Using the logarithmic evaluation described above with $T_{\max}$ determined self-consistently from the cutoff $b_{\min}(E)$, we then evaluate the
stopping power for an electron-free target; the results are summarized in Fig.~\ref{fig:E_lossBB_gene_losseft}.%

\textit{Remark on scales.}
In electron-containing conventional materials, the admissible range of transferable energy is typically limited to $T_{\max}/T_{\min}\sim10^{3}$–$10^{4}$, whereas in electron-free targets it can extend to $10^{10}$–$10^{11}$.   This broader range enters only logarithmically in the stopping term, so the familiar single-encounter mass-ratio scale $m_e/M\simeq 1/5500$ is effectively weakened to $1/1000$–$1/2000$ in our self-consistent cutoff scheme, yet the
breakeven condition remains satisfied.  Moreover, because of the logarithmic dependence, shifting the admissible range of $T$ (or equivalently $b$) by one decade
changes the total stopping by only $\mathcal{O}(10\%)$, leaving our conclusions unaffected.

\section{Energy Generation and Beam Deceleration Until Stopping}
\label{sec:lab_to_cm}

In previous sections, we evaluated energy generation and loss for infinitesimal target thicknesses at specific beam energies. Here, we examine the full process in which a beam generates energy via nuclear reactions while simultaneously losing energy, eventually coming to rest. 
When discussing energy generation, the appropriate definition of the released energy depends on how the externally supplied beam energy is treated. If the energy provided by the accelerator could be fully recovered after the reaction, it would be sufficient to consider only the expectation value of the fusion Q value. In the present scheme, however, the energy lost through stopping cannot be recovered and is irreversibly dissipated. We therefore evaluate the energy gain along the beam deceleration process itself, treating the instantaneous beam energy during slowing down as E'. 
Accordingly, the expectation value of the generated energy per reaction is taken to be $(Q+E')$, reflecting that the projectile kinetic energy is also released upon fusion.
The total energy generated during this process is expressed as:

\begin{equation}
% E_{\mathrm{gen}} = \int_{E_{\mathrm{fin}}}^{E_B} I(E)\,\rho\,\sigma(E)\,(Q+E)\,\left(\dv{E_{\mathrm{loss}}}{x}\right)^{-1}\,\mathrm{d}E,
E_{\mathrm{gen}} = \int_{E_{\mathrm{fin}}}^{E_B} I(E')\,\rho\,\sigma(E')\,(Q+E')\,
\left(\frac{\mathrm{d}E}{\mathrm{d}x}\Big|_{E'}\right)^{-1}
\,\mathrm{d}E'.
%\left(\dv{E'}{x}\right)^{-1}\,\mathrm{d}E',
 \label{eq:integral}
\end{equation}

The fusion cross section $\sigma(E)$ is shown in Fig.~\ref{fig:dt_cross_section},~\cite{DT_sigma}. The stopping power $(dE/dx)$ is the total energy-loss term discussed in Secs.~III–VI, and $I(E)$ denotes the surviving beam flux along the slowing-down trajectory (normalized so that $I(E_B)=1$). The inverse stopping power $(dE/dx)^{-1}$ converts the spatial integral over areal thickness into an integral over the beam energy.

Beam-flux attenuation is modeled by an effective attenuation cross section $\sigma_{\mathrm{eff}}(E)$ via
\begin{equation}
\frac{1}{I}\frac{dI}{dE} = -\rho \,\sigma_{\mathrm{eff}}(E)\
   \left(\frac{\mathrm{d}E}{\mathrm{d}x}\right)^{-1}\,
\end{equation}
with
\begin{equation}
\sigma_{\mathrm{eff}}(E) = \max\!\left[\,\sigma(E),\; \sigma_{\mathrm{crit}}(E)\,\right] .
\end{equation}
Here, $\sigma(E)$ denotes the nuclear reaction cross section and $\sigma_{\mathrm{crit}}(E)=\pi b_{\mathrm{crit}}^2$ quantifies beam removal by single-collision events that reduce the projectile speed below one half of its incoming value (see Sec.~VII).

Importantly, the same lower impact-parameter cutoff $b_{\min}(E)$ that was determined self-consistently for the stopping-power logarithm is also used to define the attenuation channel: events with $b<b_{\min}(E)$ are excluded from the continuous small-angle stopping and are counted as flux loss (attenuation) instead. Thus, the $b_{\min}(E)$ employed in the logarithmic evaluation of the energy-loss term and the one used to construct $\sigma_{\mathrm{eff}}(E)$ are identical by construction. This guarantees that the separation between continuous stopping and flux attenuation is applied consistently in both the loss term and the beam-survival factor $I(E)$.

\begin{figure}[h]
\centering
\includegraphics[width=0.8\textwidth]{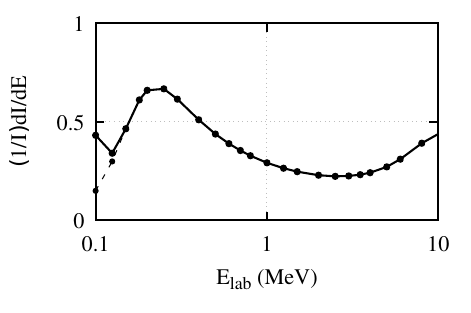}
\caption{Beam intensity decay rate as a function of beam energy (Lab frame).  The discrete points indicate values evaluated at specific beam energies, 
highlighting the crossover near $E\simeq0.13$\,MeV, while the connecting lines serve as guides to the eye.}
\label{fig:beam_decay_rate}
\end{figure}

Figure~\ref{fig:beam_decay_rate} shows the beam decay rate as a function of beam energy in the laboratory frame.  A transition is observed around 
$E \simeq 0.13$~MeV: below this energy, beam loss is dominated by velocity-changing collisions, whereas above it the attenuation is governed by nuclear reactions. In the higher-energy regime, beam attenuation is therefore controlled by fusion reactions, and the consumed beam particles are entirely converted into energy generation.  In contrast, in the lower-energy region the beam intensity is depleted primarily through nonreactive loss channels, so that the contribution to fusion energy generation is effectively extinguished.

Above $\sim0.13$~MeV, the decay rate reflects the behavior of the fusion cross section: it reaches a maximum around $E\sim0.3$~MeV and then decreases, before rising again from $\sim3$~MeV. This nonmonotonic structure originates from the competition between the energy dependence of the reaction cross section and that of the stopping power. Larger decay rates indicate a higher probability for the beam to undergo fusion at that energy, and thus a larger contribution to net energy generation.

To compute beam intensity $I(E)$, we evaluate the integral numerically, 

\begin{equation}
  I(E)=
  \exp\!\big[-\int_{E}^{E_B}\rho \,\sigma_{\mathrm{eff}}(E')\,
   \left(\frac{\mathrm{d}E}{\mathrm{d}x}\Big|_{E'}\right)^{-1}\,
   \mathrm{d}E'  \ \big]\,
%  \left(\dv{E'}{x}\right)^{-1}\,\mathrm{d}E' \ \big]\,
  \label{eq:IofE}
\end{equation}
where the initial beam intensity $I_B=1$.

In practice, the beam energy is followed down to 0.02 MeV, below which the beam is assumed to be fully stopped in the target. 
In the low-energy region, the fraction of beam particles lost through fusion reactions rapidly decreases: it amounts to about one third at 0.1 MeV and falls below the percent level at 0.05 MeV. 
Consequently, although the beam intensity continues to decrease due to nonreactive slowing-down processes, its contribution to fusion energy generation becomes negligible well below 0.1 MeV.

\begin{figure}[h]
\centering
\includegraphics[width=0.8\textwidth]{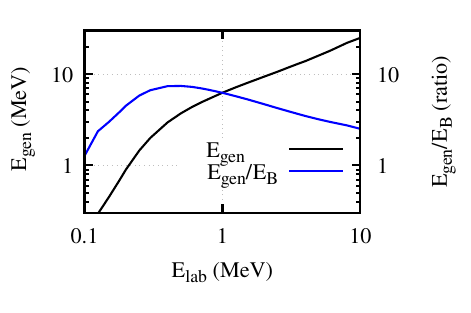}
\caption{Total energy generated as a function of beam energy.}
\label{fig:energy_generated}
\end{figure}

Next, we calculate the total energy generated using the beam intensity and cross section. 
Figure~\ref{fig:energy_generated} shows the total energy generated as a function of beam energy. 

Figure~\ref{fig:energy_generated} also shows the ratio of generated energy to input beam energy. The maximum ratio is 7.5 at around 0.4 MeV, even though only 20\% of the beam reacts before stopping. This high ratio is due to the large $Q$ value of the DT reaction.

Since electrostatic acceleration is highly efficient up to a few MeV, beam energies around 1 MeV are sufficient to maintain high energy productivity.

\section{Experimental Considerations: Acceleration Efficiency and Target Realization}

In order to approach the conditions considered in this work, an electron-free target configuration is required.
Techniques for confining charged particles using ion traps are well established, making the preparation of such targets feasible. However, practical implementation demands increased ion density. A critical factor is the reduction of electron contamination. Since the core concept of this study is to suppress energy loss by eliminating electrons—reducing it to less than one-thousandth of conventional levels—minimizing electron intrusion is essential. Ideally, electron contamination should be reduced by three to four orders of magnitude.

Even in ultra-high vacuum systems, pressures around $10^{-8}$ Pa correspond to a residual gas density of approximately $2.6 \times 10^6$ atoms/cm$^3$, implying a similar number of electrons. To achieve appreciable fusion power generation, the ion density must be significantly higher. 
For example, the ITER project~\cite{Wurzel2022Lawson, ITER_Physics_Basis_1999} is designed to operate at plasma densities on the order of
$10^{13}–10^{14}$ atoms/cm$^3$.

\subsection{Collision-Based Systems for Electron Elimination}

An alternative method to eliminate electron contamination is the use of a collider-type setup. In such systems, both the beam and the target are composed of accelerated particles, each carrying only one type of charge. This configuration inherently excludes electrons. Moreover, since both reacting species are beams, their densities can be precisely controlled, significantly enhancing the reaction rate.

In collider experiments, the reaction rate is typically expressed in terms of luminosity, which quantifies the number of interactions per unit time at the detector’s interaction point. However, for fusion reactors, reactions can occur throughout the overlapping volume of the two beams. The total reaction rate can thus be expressed as a volume integral over the region where both beams coexist.

Compared to ion trap-based targets, beam-based systems allow for greater flexibility in controlling density through beam focusing techniques. Although traditional colliders prioritize achieving the highest possible energies without considering acceleration efficiency~\cite{Energy_Efficiency}, fusion applications require careful optimization of both energy generation and efficiency. While many technical challenges remain before such systems can be realized as practical reactors, the potential merits warrant further investigation.

In beam--target configurations, the energy lost through stopping is deposited in the target and must ultimately be removed by cooling, whereas in collider-type configurations the losses arise in both beams.  While collider-type systems introduce additional challenges such as beam divergence and space-charge effects,  their consideration here serves to illustrate that Coulomb repulsion does not represent a fundamental obstacle to the breakeven conditions discussed in this work.

\subsection{Acceleration Efficiency and Effective Breakeven Condition}

In realistic systems, the beam energy must be supplied by an external accelerator, and the overall energy balance is therefore controlled by the acceleration efficiency $\varepsilon$.  Typical overall efficiencies of high-power accelerators span a wide range,
depending on the acceleration scheme and operating regime~\cite{Energy_Efficiency}.
While the local breakeven condition $R(E) > 1$ compares fusion energy generation with stopping loss
along the beam trajectory, the externally supplied electrical energy exceeds the beam kinetic energy
by a factor $\varepsilon^{-1}$. The effective condition for net energy gain is therefore
\[
R_{\rm eff}(E) = \varepsilon R(E) > 1 .
\]

An important implication of the present analysis is that the requirement on $\varepsilon$ is significantly relaxed
once electron-induced stopping is strongly suppressed.
As shown in Fig.~5, the integrated energy gain reaches
\( E_{\rm gen}/E_B \simeq 5–8 \) over a broad beam-energy range.
This implies that breakeven can be achieved even for acceleration efficiencies as low as
\[
\varepsilon_{\min} \simeq (E_{\rm gen}/E_B)^{-1} \sim 0.1–0.2 .
\]

Such values are well below the efficiencies routinely achieved by electrostatic accelerators and fall within the demonstrated range of RF-based systems.
Moreover, as discussed in Ref.~\cite{Energy_Efficiency}, further improvements in RF and power-conversion technologies remain feasible,
indicating that the proposed breakeven mechanism does not rely on near-ideal acceleration.

\subsection{Power Output}

For a general beam--target (or beam--beam) configuration, the reaction rate can be written as
\begin{equation}
\frac{dR}{dt}
=
\int \rho_B(\mathbf r)\,\rho_T(\mathbf r)\,
\sigma(E)\,v_{\mathrm rel}(E)\, dV ,
\end{equation}
where $\rho_B$ and $\rho_T$ denote the densities of the two reacting species and the integral extends over the interaction volume.
The corresponding energy-generation rate is obtained by weighting each reaction by $(Q+E)$,
\begin{equation}
\frac{dE_{\mathrm gen}}{dt}
=
\int \rho_B\,\rho_T\,
\sigma(E)\,v_{\mathrm rel}(E)\,(Q+E)\, dV .
\end{equation}
In any realistic fusion reactor configuration, achieving an appreciable reaction rate requires increasing both the particle densities $\rho_{\mathrm B,T}$ and the effective interaction volume $V$, which remain central subjects of ongoing experimental and technological research.

\section{Alternative Nuclear Fusion Reactions}

While the DT reaction is the most favorable for achieving breakeven owing to its large cross section at low energies, the present framework is not limited to DT fusion.
Other reactions can be assessed in the same manner by incorporating their corresponding cross sections, energy-loss mechanisms, and acceleration requirements.

As an example, the D--$^3$He reaction produces an alpha particle and a proton, with minimal neutron emission.
Although this reaction requires higher incident energies than DT fusion, its reduced neutron production offers potential advantages in terms of radiation management.
The breakeven condition for such alternative reactions can be evaluated within the same formalism developed here.

\section{Discussion: Advantages and Limitations of the Proposed Method}

Conventional fusion approaches seek to achieve breakeven through the control of high-temperature plasmas, which poses significant challenges in plasma heating and confinement.
In contrast, the approach considered here does not rely on bulk thermalization of the target, but instead exploits directed beam energies.

If electron-free or strongly electron-suppressed targets can be realized, the analysis indicates that breakeven conditions may be approached without invoking plasma heating.
Furthermore, the target itself may be configured as a beam, allowing for collider-type configurations.

The electron-free target is fundamentally distinct from neutral plasmas, consisting solely of positively charged ions.
Achieving high density in such systems requires suppressing Coulomb repulsion and developing appropriate confinement strategies.
While these challenges differ from those encountered in conventional fusion research, the present results motivate further exploration of such alternative approaches.

\section{Conclusion}

We have developed a self-consistent framework for evaluating breakeven conditions in beam--target nuclear fusion systems with electron-free or electron-suppressed targets.
By explicitly separating continuous energy loss from beam attenuation processes, we identified conditions under which the energy generated by fusion reactions can exceed the energy lost during beam deceleration.

The analysis shows that, when electron-induced stopping is strongly suppressed and acceleration efficiency is sufficiently high, directed beams with MeV-scale energies can satisfy the breakeven condition without relying on plasma heating.
The framework accommodates a variety of target configurations, including collider-type arrangements, and can be readily extended to alternative fusion reactions.

While significant technological challenges remain—particularly in achieving high-density electron-free targets and managing Coulomb repulsion—the present work establishes the physical principles under which net energy generation becomes possible.
These results provide a foundation for future studies exploring experimental realizations of beam-based fusion concepts.

\bibliographystyle{unsrt}
\bibliography{references}

\end{document}